# The shaft of the type 1 fimbriae regulates an external force to match the FimH catch bond

Johan Zakrisson[†, ‡], Krister Wiklund[†, ‡], Ove Axner[†, ‡], and Magnus Andersson[†, ‡]*

[†]Department of Physics, Umeå University, [‡]Umeå Centre for Microbial Research (UCMR), Umeå University, SE-901 87 Umeå, Sweden;

*Corresponding author: Magnus Andersson, Department of Physics, Umeå University,

SE-901 87 Umeå, Sweden, Tel: + 46 – 90 786 6336, FAX: +46 – 90 786 6673,

**E-mail:** magnus.andersson@physics.umu.se

**Running title:** Damping properties of type 1 fimbriae

## ABSTRACT

Type 1 fimbriae mediate adhesion of uropathogenic *Escherichia coli* (UPEC) to host cells. It has been hypothesized that fimbriae can, by their ability to uncoil under exposure to force, reduce fluid shear stress on the adhesin-receptor interaction by which the bacterium adheres to the surface. In this work we develop a model that describes how the force on the adhesin-receptor interaction of a type 1 fimbriae varies as a bacterium is affected by a time dependent fluid flow mimicking in vivo conditions. The model combines in vivo hydrodynamic conditions with previously assessed biomechanical properties of the fimbriae. Numerical methods are used to solve for the motion and adhesion force under the presence of time dependent fluid profiles. It is found that a bacterium tethered with a type 1 pilus will experience significantly reduced shear stress for moderate to high flow velocities and that the maximum stress the adhesin will experience is limited to ~120 pN, which is sufficient to activate the conformational change of the FimH adhesin into its stronger state but also lower than the force required for breaking it under rapid loading. Our model thus supports the assumption that the type 1 fimbriae shaft and the FimH adhesin-receptor interaction are optimized to each other, and that they give piliated bacteria significant advantages in rapidly changing fluidic environments.

## INTRODUCTION

Adhesion of bacteria to cells is commonly mediated via specific adhesin-receptor interactions. In vivo these bacteria are in general exposed to strong shear stress imposed by fluid flows such as expel of urine, mucosal secretions, blood, and intestinal flows that act as natural defense mechanisms against colonization of pathogenic bacteria. These fluidic flows expose bacteria to drag forces, which sometimes can be considerable, that the bacteria must sustain in order not to be flushed away. In fact, a bacterium attached to a surface of a host cell can be exposed to severe fluid drag forces with fast time dependent changes in both magnitude and directions (1–3). For example, in the urinary tract region the urine flow can reach velocities as high as 1 m/s and be turbulent (4, 5).

Uropathogenic *Escherichia coli* (UPEC) are a type of bacteria that express their adhesins on the tip of μm long helical structures called fimbriae or pili (6). It has been shown that these pili are essential for colonization of the urinary tract (7). They have a unique biomechanical behavior with a strongly non-linear extension-*vs.*-force dependence, and they are thereby believed to play an important role for the attachment of bacteria to host tissue in a fluidic environment (8). Moreover, for type 1 pili, which predominantly are expressed on bacteria in the lower urinary tract and the bladder, it has also been demonstrated that the adhesin-receptor interaction (FimH-mannose) is strongly non-linear. In fact, by the use of steered molecular dynamics simulations, flow chamber, and force spectroscopy experiments it has been shown that the FimH adhesin mediates a catch-bond, i.e., a bond whose strength is increased by a force (9, 10). It is assumed that an advantage of the catch-bond in comparison to the slip bond (for which the lifetime decreases exponentially with the force to which it is exposed) is that it gives bacteria a possibility to spread across a target surface under low shear, but still adhere strongly under exposure to high shear forces (9); bacteria adhering with catch-bonds to a surface will show rolling behavior when exposed to low shear but switch to stationary adhesion when exposed to a high shear. In addition, it is also assumed that a catch-bond is not as amenable to soluble receptor-like inhibitors as a slip bond.

Since the FimH adhesin is attached to a helical pilus, and both display a non-linear dependence on force, it is plausible that, as a means to optimize bacterial adhesion under flow (e.g., to optimize the adhesion lifetime under certain conditions), the mechanical properties of the type 1 pili shaft have coevolved with those of the FimH adhesin (11). If this is to be validated, and to assess to which degree they jointly can reinforce the ability of bacteria to sustain severe fluid drag forces, an improved understanding of the bacterial adhesion process is needed, which in turn requires detailed knowledge about the adhesin, the pili, as well as the physical interplay of bacteria with its environment.

The shaft of the type 1 pilus, onto which the tip containing the adhesin is attached, is ~ 7 nm wide and anchored to the bacterial cell. It is composed of thousands of FimA subunits that are connected in a head-to-tail manner to a chain that coils into an ordered helical arrangement. This complex structure has been investigated in detail by imaging techniques, e.g., transmission electron microscopy and atomic force microscopy, to provide conceptual models of the function in an in vivo situation (12, 13). To elucidate the mechanical properties of the shaft, force spectroscopy measurements by means of optical tweezers and atomic force microscopes have been carried out on several types of pili, not only type 1 but also P, S, F1C and type 3 (14–19). It has been found that under exposure to force, above the so-called uncoiling force, which is the force required to break turn-to-turn interactions in the coil, the shaft can be uncoiled into a linear chain whose length is several times its folded length (20). When the force varies around the uncoiling force, the type 1 pili can also, similarly to P pili (14, 15), be uncoiled and recoiled a multitude of times without any sign of fatigue. In addition, the uncoiling force has been found to depend on the extension velocity; above the so-called corner velocity is depends logarithmically on the elongation velocity (18, 21). Hence, helix-like pili exhibit dynamic properties (15).

Investigations of the adhesin, the shaft of the pili, and cell-surface adhesion have been carried out by imaging techniques, force spectroscopy, flow chamber experiments, and computer simulations (12, 16, 21–23). However, there is only a limited amount of work performed that addresses several of these entities jointly, e.g., the physical interplay between the adhesin, pili, cell, surface, and fluid. The reason is difficulties in relating the mechanical properties at a nano- and microscale to an in vivo based situation, i.e., bacterial adhesion to a surface under hydrodynamic interactions from the flow. One such example is the work by Whitfield et al. who, by simulations, examined shear-stabilized rolling of *E. coli* (23). The authors scrutinized the adherence of a cell (modeled as a spherical particle) with protruding filaments of limited flexibility and equipped with catch-bond adhesins to a surface under different shear rates, related to previous flow chamber experiments. The simulations demonstrated that shear-stabilized rolling can appear as a consequence of an increased number of low-affinity bonds caused by an increased fimbrial deformation (bending) with shear. However, the pili were only modeled as bendable structures; they lacked important biomechanical properties such as uncoiling. In addition, the simulated shear stress was significantly lower than what is expected in certain regions in vivo in which typical fluid velocities are high and flows are turbulent.

Recently, we presented a general quantitative physical model that elucidates how helix-like biopolymers behave when attached to a spherical cell under exposure to fluid shear (24). In that work both P and type 1 pili were used as model systems. The sticky-chain model (25), which relates the extension velocity of a pilus to its uncoiling force, was combined with the drag force that acts on a bacterial cell in a fluidic flow. It was found that the uncoiling ability of a pilus, which can give the bacteria a "go with the flow" ability (24), can reduce significantly the force the adhesion-receptor bond experiences to levels well below that of the fluid shear force. An example of a result from that work is shown in Fig. S1 (supplementary materials), which illustrates the dependence of the flow on the drag force for a bacterium attached to a surface with a stiff linker (*dashed dotted black line*) and a type 1 pilus (*dashed blue line*). It was found that when a bacterium attached with a type 1 pilus is exposed to a fluid flow of 15 mm/s the drag force to which the adhesin-receptor bond is exposed is reduced by 50 % as compared to when it is attached by a stiff linker (or a structure without any elastic and flexibility). It was concluded that a type 1 pilus protruding from a sphere, in the presence of a flow, and for a limited amount of time (during the uncoiling process), can act as a shock-absorber that momentarily can dampen an external force, so the load on the adhesin-receptor bond by which the bacterium adheres to the host cell will be significantly reduced (24). That result supports the previous proposed shock-absorber model presented in (11).

It was moreover argued that also other types of biopolymers, with intrinsic properties similar to those of type 1 pili, can, for certain fluid velocities and under short time scales (a few milliseconds), significantly reduce drag forces by uncoiling. It was hypothesized that this type of damping property would be beneficial for bacteria in a fluid environment of turbulent character, where rapid changes in magnitude and direction of the flow otherwise could lead to high stress on the adhesin causing detachment.

However, the fluid conditions in the "go with the flow" model (24) were the simplest possible. The model is one-dimensional; it considers a bacterium far away from any surface (i.e. it neglects surface effects) that, due to the drag force, moves solely in the direction of the flow. It was also assumed, for simplicity, that the flow was uniform and time independent. This differs considerably from the conditions to which bacteria are exposed in vivo, which consist of time dependent and turbulent shear flows. Examples of turbulent fluid environments in vivo can be found in the upper urinary tracts, where strong vortices are created from the peristaltic motion, and in the urethra during expel of urine. It has been shown that the peristaltic motion can transport boluses at mean flow velocities up to 30 mm/s (26) and that the contractions in the ureter will induce a flow field that is complex with both backflows and vortex motion (27, 28) causing high variations of shear rates. In addition, in the urethra the flow can locally reach as high as 1000 mm/s (5), which can transform a laminar flow to turbulent. It is also commonly known that turbulent flows give rise to fluid "tsunamis", sweeps, and ejections that can reach into the laminar viscous sublayer and thereby expose bacteria to significantly increased drag forces during milliseconds long periods (29, 30). With these conditions in mind we developed a model and a simulation procedure to shed light onto how piliated bacteria can be affected by such environments during their earliest attachment process. In the current work we thereby present a refined physical model that simulates a bacterium attached to a surface by a FimH adhesin mediated by a type 1 pilus that is exposed to both time-independent and time-dependent shear flows similar to those experienced in vivo.

The work thus serves the purpose of elucidating the role of pili uncoiling for the fate of a bacterium attached by a single pilus, in particular its trajectory in the closest proximity to the attachment point as well as its adhesion probability (i.e., the survival probability of the FimH adhesin). It is found that the unique extension properties of the shaft significantly reduce the load on the adhesin to a level that corresponds well to the force that previously has been experimentally and theoretically shown to activate the catch-bond to give the longest survival probability of the FimH-receptor bond (31), hence correlating some of the key properties of the pili shaft to those of the adhesin-receptor bond. The work thereby adds support to the assumption that the shaft and the adhesion bond have coevolved during the evolution process.

## THEORY AND MODEL SYSTEM

### Model system: Fluid conditions

Based upon the aforementioned in vivo conditions, for example the flow in urethra, the fluid was modeled as a pipe flow with a turbulent core and a thin boundary layer close to the surface as illustrated in Fig. S2 *A*. The flow in the layer close to the smooth surface, the viscous sublayer, in which the bacteria attach, was assumed to be laminar with a velocity profile increasing linearly from zero at the surface to a maximum value determined by the core flow.

Moreover, in the presence of turbulence it has been shown that high velocity flows occasionally sweep down from the core flow into the thin laminar sublayer, as illustrated in Fig. S2 *B* (2, 29). These sweeps are created by strong vortices outside the viscous sublayer, temporarily increasing the shear rate near the surface (32). The effect of these temporary sweeps was modeled by momentarily changing the shear rate of the velocity profile in the viscous sublayer as is illustrated by the two flow profiles in Fig. S2, *A* and *B*.

A nominal shear rate, $S_0$, was defined, chosen to correspond to a reference flow velocity of 8 mm/s at a distance of 3 μm above the surface, which represents a shear rate of around 2700 $s^{-1}$. This reference velocity was estimated by considering the general flow in a urethra, i.e., a 5 mm diameter pipe with a center velocity of 1000 mm/s, where the laminar sublayer close to surface was modeled in a standard fashion using a linear velocity profile determined by the pipe wall shear stress (30).

To mimic flows and their changes in our model, simulations were made at three different shear rates, $S_1$, $S_2$, and $S_3$, representing values of $S_0$, $2S_0$, and $3S_0$, respectively, where the two latter predominantly represent an increased rate during a momentary sweep. A transient increase of the flow implies a sudden increase of the drag force adding tensile stress to the pilus and thereby to the adhesin.

## Model system: Tethered bacterium exposed to fluid flow

Figure 1 displays a schematic illustration of the model presented in this work. A bacterium is attached to a surface by a pilus and exposed to a linear flow profile. The model takes into account four important physiological conditions; *i)* a 2D-motion of bacteria relative the surface, *ii)* time dependent fluid velocity shear profiles, *iii)* drag force corrections for near surface motion, both parallel (33) and perpendicular motion (34), and *iv)* the allosteric catch bond properties of the FimH adhesin.

Figure 1 here.

## Pili uncoiling under 1D conditions

The model developed here is based in part on our previous one-dimensional model for damping of the adhesion force of a bacterium adhering by a single pilus exposed to a flow (24). That work describes a bacterium modeled as a rigid sphere in a uniform flow attached to a surface with a pilus of length $L$. The sphere was constrained to move in one dimension, parallel to the flow, exposed to a Stoke drag force given by,

$$F_{\mathrm{D}} = 6\pi\eta r(v_{\mathrm{f}} - v_{\mathrm{b}}),$$ (1)

where $\eta$ is the dynamic viscosity of the fluid, $r$ is the radius of the sphere, $v_{\mathrm{f}}$ is the fluid velocity, and $v_{\mathrm{b}}$ is the velocity of the bacterium (the latter two with respect to the surface). The sphere was assumed to be neutrally buoyant and to follow the flow when it suddenly got attached by a pilus (alternatively, when it suddenly got exposed to a flow). Since the sphere is anchored by an extendable pilus, its velocity is equal to the elongation velocity of the pilus, i.e., $v_{\mathrm{b}} = \dot{L}$, which in turn is related to the tensile force to which the pilus is exposed (which is equal to that applied to the sphere as well as that experienced by the adhesin), $F_{\mathrm{P}}$, that can be written as [see (24) for a detailed derivation],

$$\dot{L}(F_{\mathrm{P}}) = \dot{L}^{*} e^{(F_{\mathrm{P}} - F_{\mathrm{SS}})\Delta x_{\mathrm{AT}}\beta} \left[ 1 - e^{-(F_{\mathrm{P}} - F_{\mathrm{SS}})\Delta x_{\mathrm{AB}}\beta} \right],$$ (2)

where $\dot{L}^{*}$ is the corner velocity for the uncoiling of the pilus, $F_{SS}$ is the steady state uncoiling force, $\Delta x_{\mathrm{AT}}$ is the turn-to-turn bond length from the ground state to the transition state , $\beta = 1/kT$, where $k$ is the Boltzmann's constant and $T$ is the temperature, and $\Delta x_{\mathrm{AB}} = \Delta x_{\mathrm{AT}} + \Delta x_{\mathrm{TB}}$, where $\Delta x_{\mathrm{TB}}$ is the distance from the transition state to the open state. Due to the low mass of the bacterium the effect of inertia was neglected, since it gives rise to forces several magnitudes lower than the fluid forces. Moreover, since and the model was one-dimensional, $F_{\mathrm{P}}$ was at every instance considered equal to $F_{\mathrm{D}}$ . This model demonstrated that the force to which an adhesin is exposed can be significantly reduced due to the elongation properties of the pilus; as long as the pilus is extending the force acting on the adhesin is lower (often significantly lower) than that of a stiff linker [for which the force is given by Eq. (1) with $v_b$ being zero]. Although this one dimensional model can illustrate the ability of the pilus to dampen the drag force, it cannot provide a full picture of the adhesion phenomena of a bacterium in a flow. In particular, it does not take the shear profile and near-surface effects into account. In order to investigate this, a 2-dimensional (2D) model needs to be considered.

## 2D-movement of a bacterium attached by a pilus under a linear shear profile

In our refined model, presented here, a tethered bacterium (again modeled as a sphere) is allowed to move in two dimensions in a fluid with a linear shear profile, i.e., with $\mathbf{v}_f(y) = Sy\hat{\mathbf{x}}$ where $S$ is the shear rate, $y$ is the distance from surface (measured in the direction normal to the surface), and $\hat{\mathbf{x}}$ is the unit vector parallel to the surface (see Fig. 1). Since a bacterium attached to the surface by a pilus and exposed to a force will move in the flow it will in general have a position dependent velocity that can be expressed as

$$\mathbf{v}_b(y) = v_{bx}(y)\hat{\mathbf{x}} + v_{by}(y)\hat{\mathbf{y}},$$ (3)

where $v_{bx}(y)$ and $v_{by}(y)$ are the components of the velocity in the $\hat{\mathbf{x}}$- and $\hat{\mathbf{y}}$-directions, respectively, and where $\hat{\mathbf{y}}$ is the unit vector perpendicular to the surface (se Fig. 1). Note that while $v_{bx}(y)$ is positive, $v_{by}(y)$ takes a negative value as the bacterium approaches the surface. In addition, the model can also allow for the influence of sweeps by allowing the shear rate to be time dependent, i.e., $S(t)$. By assuming that $y$ describes the position of the sphere, which in turn is time dependent, the bacterium will experience a flow velocity, $\mathbf{v}_f$, that depends on both its position, $y(t)$, and time [through $S(t)$], i.e.,

$$\mathbf{v}_f\left[y(t),t\right] = S(t)y(t)\hat{\mathbf{x}}.$$ (4)

Hence, the fluid model cannot only take into account the altered flow velocity experienced by the sphere as it moves in the fluid, it can also depict the influence of a change in flow velocity on the sphere (due to a time-dependent shear rate).

A bacterium exposed to a flow will be affected by several forces of which the two dominant, as is illustrated in Fig. 1, are the pilus force, $\mathbf{F_P}$, which has a magnitude of $F_P$ and is directed along the direction of the pili, here denoted by $\hat{\mathbf{r}}$, and the fluid drag force, $\mathbf{F_D}$, originating from the relative motion of the bacterium with respect to the fluid, which conveniently can be written in terms of their x- and y-components as

$$\mathbf{F}_{\mathrm{P}}(t) = F_{\mathrm{P}}(t)\hat{\mathbf{r}} = F_{\mathrm{Px}}(t)\hat{\mathbf{x}} + F_{\mathrm{Py}}(t)\hat{\mathbf{y}}, \tag{5}$$

$$\mathbf{F}_{\mathrm{D}}(t) = F_{\mathrm{Dx}}(t)\hat{\mathbf{x}} + F_{\mathrm{Dy}}(t)\hat{\mathbf{y}} . \tag{6}$$

Lift forces arising from shear flow (35, 36) and from the presence of a wall (37) can be neglected since these are insignificant compared to the other forces. As for our one-dimensional model (24), the sphere is assumed to be neutrally buoyant over the time scale studied.

As was discussed in Ref. (24), the drag force experienced by the sphere will depend on the relative motion of the bacterium with respect to the velocity of the fluid. However, since the sphere is close to a wall, wall proximity effects need to be taken into account. This is normally done by modifying the Stoke drag force by the use of a "correction term". However, since the motion of the sphere has components both in the parallel and perpendicular directions with respect to the surface, and the effect of the presence of the wall is dissimilar in the two directions, it is appropriate to write (34, 38)

$$F_{\mathrm{Dx}}\left[ y(t),t \right] = C_{\mathrm{x}}\left[ y(t) \right] 6\pi\eta r \left[ S(t)y(t) - v_{\mathrm{bx}}(t) \right], \tag{7}$$

$$F_{\mathrm{Dy}}\left[ y(t),t \right] = -C_{\mathrm{y}}\left[ y(t) \right] 6\pi\eta r v_{\mathrm{by}}(t), \tag{8}$$

where $C_{\mathrm{x}}[y(t)]$ and $C_{\mathrm{y}}[y(t)]$ are corrections to the Stoke drag force experienced by a bacterium close to a wall moving in the x- and y-directions, respectively. For motions parallel to the surface we use the 5$^{\mathrm{th}}$ order correction factor derived by Faxen, e.g., given by Eq. 7-4.28 in reference (34), which reads

$$C_{\mathrm{x}}\left[ y(t) \right] = \left\{ 1 - \frac{9}{16}\left[ \frac{r}{y(t)} \right] + \frac{1}{8}\left[ \frac{r}{y(t)} \right]^3 - \frac{45}{256}\left[ \frac{r}{y(t)} \right]^4 - \frac{1}{16}\left[ \frac{r}{y(t)} \right]^5 \right\}^{-1}, \tag{9}$$

where $r$ is the radius of the bacterium and $y$ is the distance from surface to the center of bacterium. To correct for the motion perpendicular to the surface we use the approximated correction factor that is valid for $y/r > 1.1$, which can be written as (39)

$$C_{\mathrm{y}}\left[ y(t) \right] = \left\{ 1 - \frac{9}{8}\left[ \frac{r}{y(t)} \right] + \frac{1}{2}\left[ \frac{r}{y(t)} \right]^3 - \frac{57}{100}\left[ \frac{r}{y(t)} \right]^4 + \frac{1}{5}\left[ \frac{r}{y(t)} \right]^5 + \frac{7}{200}\left[ \frac{r}{y(t)} \right]^{11} - \frac{1}{25}\left[ \frac{r}{y(t)} \right]^{12} \right\}^{-1}$$

.
$$\tag{10}$$

Since the effect of inertia can be neglected, force balance must prevail at every moment. This implies that

$$F_{Dx}\left[y(t),t\right]=F_{Px}\left[y(t),t\right]=F_P\left[y(t),t\right]\cos\left[\theta(t)\right], \qquad (11)$$

$$F_{Dy}\left[y(t),t\right]=F_{Py}\left[y(t),t\right]=F_P\left[y(t),t\right]\sin\left[\theta(t)\right], \qquad (12)$$

where $\theta(t)$ is the angle of the pilus with respect to the surface as indicated in Fig. 1.

These equations need to be solved for the position and velocity of the sphere, as well as the drag and pili forces, under the condition that the movement of the sphere is constrained by a tether behaving as a pilus according to the model given above [see Eq. (2)]. However, the coupling of the position and the velocity of the sphere to the instantaneous drag and pili forces, i.e., the Eqs (7) – (12), and that of the pilus force to the elongation velocity of the pilus, Eq. (2), make the system difficult to solve by analytical means. In short, they imply that as the sphere is exposed to a flow and moves towards the surface, it will experience both a position dependent velocity and be exposed to a position dependent drag force (which both additionally are time dependent in the presence of sweeps), which in turn give rise to a varying force to which the pilus is exposed. Instead, as described in some detail in the *supplementary material*, the set of equations given above has been solved by numerical means.

As is discussed in some detail below, and as is schematically illustrated in Fig. 2, this system of equations indicates that the sphere will move in curved trajectories. For low flows (and thereby low drag forces), the pilus will not uncoil, whereby its length, $L$, will not change, being equal to its original coiled length, $L_0$. The sphere will then be constrained into a circular motion controlled by a pilus with a fixed length $L_0$, as is illustrated by the dashed line in Fig. 2.

Figure 2 here.

For high enough flows, however, the force acting on the pilus will induce uncoiling (and possibly also subsequent recoiling), which gives rise to an extension (and a possible contraction) of the pilus, i.e., a time dependent length, $L(t)$, whose elongation velocity at each moment in time is given by Eq. (2). As is further alluded to in the supplementary material, the part of the bacterium velocity due to unfolding will be in the direction of the external force, i.e., in the $\hat{\mathbf{x}}$-direction. This will not only affect the motion trajectory of the bacterium, as is illustrated by the solid arrows in Fig. 2, it will also decrease the force the adhesin experiences in a manner similar to that of the one-dimensional model described in Ref. (24).

In order to investigate the influence of the extendibility of the pilus on the model system — in particular the movement of the bacterium, the force acting on the adhesin, and the survival probability of the bacterium — two types of tethers that constrain the motion of the sphere were considered; a stiff linker and an extendable pilus with biomechanical properties of the type 1 pilus. The motion of a bacterium attached by a stiff linker is therefore used as a reference system to which we compare the biomechanical properties of an extendable pilus with dynamic properties.

### The FimH-mannose bond

The type 1 pili adhesin, FimH, is expressed at the tip of the pilus. It anchors the bacteria to receptors (mannose) on the surface of the host and hence mediates the forces that act on the bacterium. The FimH-mannose bond has successfully been modeled by an allosteric catch-bond model that can predict the lifetime of a bond that can switch between three different states, a weak, a strong, and an unbound state (22). The model explains and predicts *E. coli* bacteria rolling on mannose-BSA and force spectroscopy data performed by constant-velocity AFM (22, 31).

The probability as a function of time of the FimH-mannose bond to reside in state $i$, $B_i$, where $i$ being 1 or 2 represents the weak and strong bound states, respectively, can be described by two differential equations as (22)

$$\frac{dB_1(t)}{dt} = -\left\{k_{10}\left[F(t)\right] + k_{12}\left[F(t)\right]\right\}B_1(t) + k_{21}\left[F(t)\right]B_2(t),$$  (13)

$$\frac{dB_2(t)}{dt} = k_{12}\left[F(t)\right]B_1(t) - \left\{k_{20}\left[F(t)\right] + k_{21}\left[F(t)\right]\right\}B_2(t),$$  (14)

where $k_{10}, k_{20}, k_{12},$ and $k_{21}$ represent state transition rate coefficients, with an exponential dependence on the force, described by the Bell equation, $k_{ij}[F(t)] = k_{ij}^o \exp\left[F(t) \cdot x_{ij}/kT\right]$, where $k_{ij}^o$ is the thermal rate constant, $F$ represents the applied force, $x_{ij}$ is the distance to the transition state, and $kT$ is the thermal energy (40). These two equations determine the FimH-mannose bond survival probability, $B_1$ and $B_2$, as a function of time after bond formation for a given force.

In our simulation of the full fluid-bacteria-pili-FimH system [described by the Eqs (2), (7), (8), (13), and (14)] we used the transition rate parameter values given in Supplementary Table S1 [which originate from the Refs (31)]. As is described in some detailed in the supplementary material, first the force responses of the two tethers, stiff linker and extendable pilus, were determined by solving the Eqs (2), (7), and (8). The time-dependent force data from such simulations were then used to assess the FimH-mannose bond survival probability as a function of time by solving the Eqs (13) and (14).

## RESULTS

### Trajectory for an attached bacterium and the force experienced by the adhesin in a steady flow

To investigate the 2D movement trajectory of a bacterial cell during attachment to a surface exposed to different flow conditions, three shear rates, $S_1$, $S_2$, and $S_3$, were considered in the simulations. The bacterium was assumed to have an effective radius of 1.0 µm and to be equipped with a single 2 µm long tether. It was initially positioned (t = 0) directly above the anchoring point (i.e., at x = 0, y = 3 µm). The flow exposed the bacterium to a drag force [given by the Eqs (7) and (8)] that sets the bacterium into motion.

The tether was first given properties of a stiff linker (i.e., with no elasticity). For this case, the cell moved along a circular trajectory, as is shown by the solid black line in Fig. 3 *A*. As is alluded to in the supplementary materials section, the simulations were terminated when the membrane of the bacterium was at a distance 0.2 µm above the surface (termed the simulation termination distance, STD). This was caused by the fact that the Stoke's drag force correction coefficients are not valid for distances close to the surface.

Figure 3 here.

The stiff linker was then replaced with an extendable pilus (with biomechanical properties given by those of type 1 pili). Since its elongation velocity, modeled by Eq. (2), depends on the drag force, it is directly related to the shear rate, and thereby it varies with position above the surface. At high shear rates the drag force applies sufficient tension to the pilus to uncoil it. Following previous experimental findings it was assumed that the uncoiling commences at a force of 30 pN (16). For the case with the lowest shear rate, $S_1$, and as is shown by the green dashed line in Fig. 3 *A*, this force is obtained at a position of around 2 µm after which the bacterium takes a slightly shallower path in comparison to the stiff linker.

For the two higher shear rates, $S_2$ and $S_3$, the trajectories become significantly elongated. Initially, they coincide with that of the stiff linker until the tensile force reaches the threshold value for uncoiling. In the case with a shear rate of $S_2$ (*blue dashed dotted curve* in Fig. 3 *A*), the bacterium is translated to a position 11 µm from the tether point before it reaches the STD. In the supplementary material, movie S1 shows the movement of the bacterium at a shear rate of $S_2$ (both for a stiff linker and a type 1 pilus). For the case with a shear flow of $S_3$ (*red short dashed curve* in Fig. 3 *A*), the bacterium uncoils completely before it reaches the STD, which it does at a position 14 µm from the tether point. From this position, it acts as a stiff linker.

Figure 3 $B$ shows the corresponding force acting on the pilus (thereby also on the adhesin-receptor bond at the anchor point) as a function of time during the trajectories explored by the bacterium in Fig. 3 $A$. As above, the solid curves represent the stiff linker while the various dashed curves correspond to the type 1 pilus, both types exposed to the three different shear rates (where the green, the blue, and the red curves correspond to shear rates of $S_1$, $S_2$, and $S_3$, respectively). The data shows that even though the trajectories for the bacterium tethered with a stiff linker (*solid curves*) are identical for all shear rates (all represented by the black solid curve in Fig. 3 $A$), the forces on the anchor point differ significantly. The peak forces for the stiff linker exposed to the three shear rates are ~100, ~225, and ~350 pN, respectively.

A type 1 pilus shows for the lowest shear force, $S_1$, a similar response to that of the stiff linker (*the green curves*) since it solely experiences a low uncoiling velocity. The simulations indicate that, for the higher shear rates, the elongation properties of the pilus give rise to a significant reduction of the peak force. For both the $S_2$ and $S_3$ shear rates (*the blue and the red dashed curves*) the force is reduced to ~120 pN, which is a factor 2 and 3 lower than for the case when the bacterium is attached with a stiff linker. In addition, the red curve shows that for the highest shear rate, $S_3$, the force eventually increases from 120 pN to 350 pN at ~ 1.2 ms. This sudden increase in force originates from the fact that the pilus has been fully uncoiled and then acts as stiff linker.

## Trajectory for an attached bacterium and the force experienced by the adhesin in a transient flow

The fluid dynamics of the urinary tract system is complex with peristaltic activities as well as high fluid flow velocities leading to turbulence. Turbulence, in turn, gives rise to transient sweeps and ejections that instantaneously change the fluid profile in the viscous sublayer. To better understand how turbulence can affect a tethered bacterium, sweeps were simulated by adding a transient increase of the shear flow at given onset times; 0, 0.25, 0.5, 0.75, and 1 ms.

Figure 4 $A$ and $B$ show simulations of sweeps (modeled as step increases in the shear rate from $S_1$ to $S_2$, and from $S_1$ to $S_3$, respectively) that appear at different onset times. As is shown by the solid curve in Fig. 4 $A$, for a bacterium attached with a stiff linker the trajectories are again identical (they all overlap). For an extendable pilus, exposed to a sudden sweep with a shear rate of $S_2$, the total translation length decreases significantly (from 11 to 4.5 µm) as the onset of the sweep is increased (from 0 to 1 ms). This comes from the fact that for the late onset times the bacterium is exposed to a force above the uncoiling force for only a fraction of its trajectory. Hence it reaches the STD mainly coiled. As is illustrated in Fig. 4 $B$, for the largest shear rate, $S_3$, the pilus reaches full elongation (14 µm) for all onset times up to 0.5 ms.

Figure 4 here.

The force acting on the anchor point as a function of time for three of the cases with sudden onset of sweeps considered above is shown in Fig. 5, where the two panels *A* and *B* represent sweeps with shear rates of $S_2$ and $S_3$, respectively. The solid curves show the force acting on the anchor point for a stiff linker for different onset times. The red solid curve corresponds to an onset time of 0 ms, while the green and blue curves represent onset times of 0.5 and 1 ms, respectively. The simulations show that the onset of the sweep instantaneously increases the force. The green and the blue solid curves in Fig. 5 *A* show that for the lower sweep shear rate, $S_2$, the force increases from 100 to ~225 pN, while Fig. 5 *B* indicates that for the highest sweep shear rate, $S_3$, it increases up to ~350 pN.

Figure 5 here.

The dashed curves in Fig. 5 represent a bacterium attaching by a type 1 pilus under the same conditions. The simulations displayed in Fig. 5 *A* representing the lower sweep shear rate, $S_2$, show that the uncoiling capability of the type 1 pilus reduces the force on the adhesin (as compared to the stiff linker) from ~225 to ~120 pN irrespectively of the onset time of the sweep. For the case with the higher sweep shear rate, $S_3$, displayed in Fig. 5 *B*, the force is reduced to the same level as long as the pilus uncoils. The steep increases in force to ~350 pN at 1.2 and 2.1 ms for the cases with 0 and 0.5 ms onset time (*red and green dashed curves*), respectively, originate from the fact that the pilus has been fully uncoiled, whereby it thereafter acts as a stiff linker.

## The influence of pili uncoiling and bacterial trajectory on the survival probability of the FimH adhesin

The results presented above describe how the force acting on the anchor point (and thereby the adhesin-receptor bond) is affected by the properties of the tether and the fluid conditions. To get a better understanding of how these forces affect the lifetime of the adhesin-receptor bond, and thereby the bacterial adhesion lifetime, the model was extended to also include the catch-bond properties of the FimH adhesin (9). The adhesin was modeled as an allosteric catch-bond described by Eqs (13) and (14), with the four force-dependent kinetic coefficients used in ref. (31). In that work, the lifetime of the FimH-mannose bond was investigated for a constant loading rate (i.e. a linearly increasing force with time). In our case, we couple the FimH-mannose bond model to the actual time-dependent force to which the adhesion-receptor bond is exposed under the attachment process simulated above. As the flow exposes the bacterium to a drag force the tension on the tether changes the force on the adhesin that in turn modulates the probability of staying bound.

To verify our numerical simulations of the FimH-mannose bond model, the Eqs (13) and (14) were solved for the three loading rates previously considered by (31). The results, which are presented in Fig. S4 in the supplementary material, agree with ref. (31), confirming the numerical procedure used for solving these equations.

The probability for a bacterium to stay attached is given by the survival probability distribution of the adhesion-receptor bond. Figure 6 shows the probability distribution for a stiff linker and type 1 pilus (*black solid and blue dashed curve*, respectively) for the shear rate $S_2$ (corresponding to the blue curves in Fig. 3 *B* that initially increase linearly and thereafter flattens out). The figure shows that for a bacterium attached by a stiff linker the survival probability drops down drastically (to values close to zero) after a short time (less than 0.2 ms). For a bacterium adhering with a type 1 pilus, on the other hand, it remains high (>0.99) throughout the entire simulation (until the bacterium reaches the STD). The figure also shows, by the inset, that at a given short time (around 0.05 ms) the time development of the survival probability for both linkers changes to a plateau-like dependence. This is attributed to the fact that the bond alters conformation state, from the weak to the strong binding configuration, for which the lifetime is long for the pertinent force.

Survival probability distributions for a variety of transient flows were also simulated. Figure 6 shows such distributions for the six cases displayed in Fig. 3 *B*. Although not explicitly shown here, it was found that the situations presented in Fig. 5 representing delayed onsets of the sweep (the green and blue curves in the two panels, respectively) gave rise to survival probability distributions that are of similar form to those in Fig. 6, mainly shifted in time. The survival probability distributions presented in Fig. 6 can therefore be seen as typical for a broad variety of situations.

Figure 6 here.

## Influence of the parallel and normal drag correction coefficients on a tethered bacterium under flow

In addition to the above results we also investigated the impact of the normal and parallel drag correction coefficients [given by the Eqs (9) and (10)] in the simulations. Using the $S_2$ shear rate under the same conditions as presented in Fig. 4 *B*, four situations were compared; with and without each of the two correction coefficients. The results are presented in Fig. S5, which shows the force on the adhesin for the stiff linker and the type 1 pilus (*solid and dashed curve*, respectively) for the four cases. For the stiff linker, and using the model used in this work, which incorporates both correction coefficients, the maximum force was found to be ~225 pN. For the two cases where only one of the two coefficients was included, either the normal or the parallel, the maximum force was found to be ~190 pN. For the uncorrected case, it became ~125 pN. Thus, for a stiff tether, it was concluded that if the correction coefficients are not used the simulations underestimate the force by almost 50 %. For the case with type 1 pilus, the influence of the correction coefficients on the maximum force is significantly smaller; the maximum forces were found to be close to ~120 pN for all four cases. The reason for this is the force damping ability of the pilus, which disguises the effect of the correction coefficients. However, the time evolution of the motion, and thereby also of the force exposure, are affected by the correction coefficients. The simulations predict that the time to reach STD is five times longer when including the correction coefficients than when not.

## DISCUSSION & CONCLUSION

It is non-trivial to scrutinize experimentally the initial attachment or the adhesion process of a bacterium to a surface under different fluidic conditions. However, it is possible to characterize separate parts of the otherwise complicated cell-surface adhesion process, i.e., the receptor-ligand interaction, various properties of the pili, and the fluid-cell shear force. Receptor-ligand interactions and the biomechanical properties of a pilus can be assessed in great detail by force spectroscopic investigations in which entities such as bond lengths, bond energies, and transition rates can be assessed. These experiments, however, do not give any information of how the fluid gives rise to a shear force that acts on a tethered cell. Complementary techniques, e.g., parallel plate flow chamber instrumentation can provide information of how the drag force of a fluid interacts with a cell under various conditions, e.g. when the cell is moving freely with the fluid, rolling on a surface, or staying bound for different fluid flow velocities. However, such a technique cannot be used to scrutinize how pili interact with a surface or assess the number of pili by which a bacterium binds at each moment. Nor is it trivial to experimentally assess how intrinsic mechanical properties, such as the pili uncoiling force and bond transition rates, affect the motion of a cell in a flow and the associated force experienced by the adhesin-receptor bond. When a complex combined phenomenon is difficult to assess experimentally, computer simulations, based on appropriate models and input parameters, can be a powerful complement.

We present in this work a model that describes how the force on the adhesin-receptor bond of a type 1 pilus by which an *E. coli* binds to a surface is affected, and to which extent the bacterium can remain bound to the surface, when it is exposed to a time dependent fluid flow. It is based upon models for cell-fluid interactions, pili uncoiling, and the allosteric FimH catch-bond that previously have shown good agreement with experimental data (22, 23, 39, 41). It makes use of input parameters that have been assessed by force spectroscopy and flow chamber experiments. The model is investigated and assessed by simulations.

These simulations allow the 2D-trajectory of a tethered bacterium and the load on the adhesin-receptor bond during the translation to be followed in detail. It is found that pili uncoiling allows the bacterium to translate significant distances (~10-14 µm) in the direction of the flow while still being attached to a receptor under sustainable tension. This allows an attached bacterium to drift close to the surface where it is exposed to lower drag forces, which can increase the probability for additional pili to attach. As previously has been shown both experimentally and by simulations, the latter can result in shear-stabilized rolling, i.e., enhanced adhesion due to the fluid forces. For example, simulations aimed at describing the shear-stabilized rolling behavior of *E. coli* have showed that even though pili were simulated without uncoiling ability and the shear rates were significantly smaller than in our work, shear stress can give rise to compressive forces on the fimbriae that result in fimbrial deformation (23). Fimbrial deformation allows also short pili to attach to the surface, increasing the adhesion life time of a bacterium. Our simulations (providing 15-20 pN at $S_2$) support the previous findings (23) that the force pushing the cell towards the surface is sufficiently high to bend and compress type 1 fimbriae.

Our approach of calculating the drag force as a function of the position of the bacterium relative to the surface, while implementing the unique biomechanical properties of type 1 pili, has allowed us to estimate the survival probability of the FimH adhesin-receptor bond under dynamic flow conditions. Since the survival probability of a bond is directly related to the force to which it is exposed, our approach provides information of how the bacterial adhesion time is influenced by phenomena such as pilus uncoiling. It was found by the simulations that the maximum force experienced by the adhesin is strongly regulated by pilus uncoiling. For a shear rate of $S_2$, for which the maximum force mediated by a stiff linker was found to be ~220 pN, the uncoiling of the type 1 pilus reduces this to ~120 pN.

A previous study showed that for most effective binding of the FimH adhesin the force should be sufficient to fast enough activate the conformational change of the FimH adhesin into its stronger bound state, here denoted the critical force (31). For low forces the bond remains in the weakly bound state and has therefore a short survival probability. The force should, however, not be increased substantially above the critical force since this could overpower the bond. Once the bond has switched to its stronger conformation it will remain there for a wide range of forces (31). Hence, a structure that can modulate the force on the adhesin so it rapidly reaches the critical force, and thereafter keeps it at a sustainable level, provides the longest life time. Our work shows that the shaft of the type 1 pilus can reduce the force to a value similar to that of the critical force of the FimH (~100 pN) even if the bacteria are exposed to very high flows and thereby prolong the adhesion lifetime. The survival probability of bacteria attaching by a type 1 pilus can therefore remain high (>0.99) over a significant time. In addition, simulations concerned with various sweep onset delays (whose responses are similar to that displayed in Fig. 6 but not explicitly shown) indicate that short sweeps do not have to be devastating for the bacterial adhesion; the extensibility of pili can regulate the force during temporary sweeps to a sustainable level. This implies that a bacterium can benefit from a large survival probability even under short but intense sweeps (as long as the pili can uncoil). This implies that the bacterium can stay bound to a surface for a long time even though being exposed to a significant fluid drag force. This finding can explain why some bacteria can withstand the natural defense mechanisms of urine in vivo or the peristaltic transport of the lumen.

Many works dealing with adhesive experiments and simulations in flows neglect the influence of the correction coefficients for near-surface motion. We show here that the influence of these on the drag forces as well as the time to reach the surface is of importance for a cell tethered to a surface with a stiff linker. Both the drag forces and the times were significantly different for the cases when the coefficients were and were not included. However, the differences were found to be not as significant for the type 1 pilus. For a stiff linker the simulated forces differ by ~60 %, while for the type 1 pilus no such difference appears. The reason for the latter can be attributed to the force modulating property of the pilus, which temporarily regulates the force to down to the same value, ~120 pN, irrespectively of the actual load. For both tethers, also the time needed to reach the STD was affected by the use of the correction coefficients. The influence of these coefficients will presumably be of importance when simulating processes such as the rolling behavior of cells that have multiple bonds interacting with a surface.

Most adhesive biological processes are exposed to mechanical stress from fluid-induced shear in dynamic environments. Handling this stress is important both for bacteria initially attaching to host tissue and for those that already have colonized. For this purpose uropathogenic bacteria have developed helix-like surface organelles as sophisticated mediators of attachment. It has been demonstrated that such bacteria cannot adhere in the absence of pili (6). The same is valid for bacteria in other environments. For example, a haemagglutination assay showed that CFA/I pili expressed by enterotoxigenic *E. coli* bacteria exposed to shear flow did not stay bound when the helical region of the pili was mutated into an open coiled stalk (42). In addition, a force spectroscopy study illustrated that the CFA/I pili show similar uncoiling abilities as type 1 pili (43). All this implies that the helix-like structure of pili is important for many bacteria in the adhesion process. Moreover, since the attachment organelles always act jointly with the adhesin bond it supports the assumption previously proposed in (11) that they have co-evolved to reinforce their total influence on the survival probability of the bacterium. The present study, which demonstrates that the shaft of the type 1 pilus has an ability to regulate forces from hundreds of pN to 120 pN, which coincides with the specific properties of the FimH adhesin-receptor bond, has thus demonstrated that the assumption that the pili and the adhesin have co-evolved is reasonable.

## ACKNOWLEDGEMENT

This work was performed within the Umeå Centre for Microbial Research (UCMR) Linnaeus Program supported from Umeå University and the Swedish Research Council (349-2007-8673). It was also supported by a Young Researcher Award (*swe.* Karriärbidrag) from Umeå University (to M.A.) and by the Swedish Research Council (to O.A, 621-2008-3280).

## FIGURE LEGENDS

**Figure 1.** Symbolic representation of the nomenclature used in the model. The red point represents both the origin of the coordinate system and the position at which the adhesin adheres to a surface receptor. All entities displayed in the figure ($\theta$, $F_P$, $F_{P_x}$, $F_D$ etc.) take generally positive values and are described in detail in the text.

**Figure 2.** The trajectory of a bacterium tethered to a surface with a stiff linker (*black dashed arrow*) and an extendable pilus that elongates by uncoiling when the shear force is higher than the uncoiling force (*gray solid arrows*).

**Figure 3.** Panel A: The trajectory of a bacterium tethered with either a stiff linker (*solid black curve*) or a type 1 pilus (*dashed curves*) exposed to different flows, $S_1$, $S_2$, and $S_3$. The bacterium modeled has an effective radius of 1.0 µm and the length of the tether is 2 µm. The position given represents the center position of the bacterium. The gray dashed horizontal line represents therefore the position at which the bacterium touches the surface. Panel B: The force experienced by the adhesin in the various cases.

**Figure 4.** The trajectory of a bacterium exposed to five different transient sweeps. The solid black curves correspond to a bacterium attached by a stiff linker while the dashed curves represent the trajectories for the bacterium attached with a type 1 pilus. In all cases, at time 0 the bacterium was exposed to a flow profile with the nominal shear rate of $S_0$ (corresponding to a linear shear with a flow velocity of 8 mm/s 3 µm above the surface). The sudden sweeps increase the shear rate at given times, as indicated in the figure legends (after 0, 0.25, 0.5, 0.75, and 1 ms). Panel A: The shear rate was increased by a sweep from $S_0$ to $S_2$. Panel B: The shear rate was increased by a sweep from $S_0$ to $S_3$.

**Figure 5.** The force experienced by the adhesin for three of the flow conditions considered in Fig. 4 (when the transient sweep increases the shear rate after 0, 0.5, and 1 ms). The denotations of the various curves and the two panels are the same as in Fig. 4.

**Figure 6.** The survival probability for a FimH-mannonse bond when the bacterium is tethered via a stiff linker (*solid curves*) or a type 1 pilus (*dashed curves*) and exposed to three different shear rates, $S_1$, $S_2$, and $S_3$. The figure shows that for the stiff linker the survival probability rapidly decreases towards zero in less than 0.2 ms whereas for the type 1 pili it remains close to 1 for a considerable time. The inset illustrates the decrease of the survival probability of the bacteria attached by the two types of tether. After 1.2 ms at $S_3$ the pili is fully uncoiled (*red dashed curve*), i.e. the survival probability will rapidly go towards zero (not explicitly shown). The plateau represents when the catch bond has changed conformation state, from the weak to the strong binding configuration.

**FIGURES**

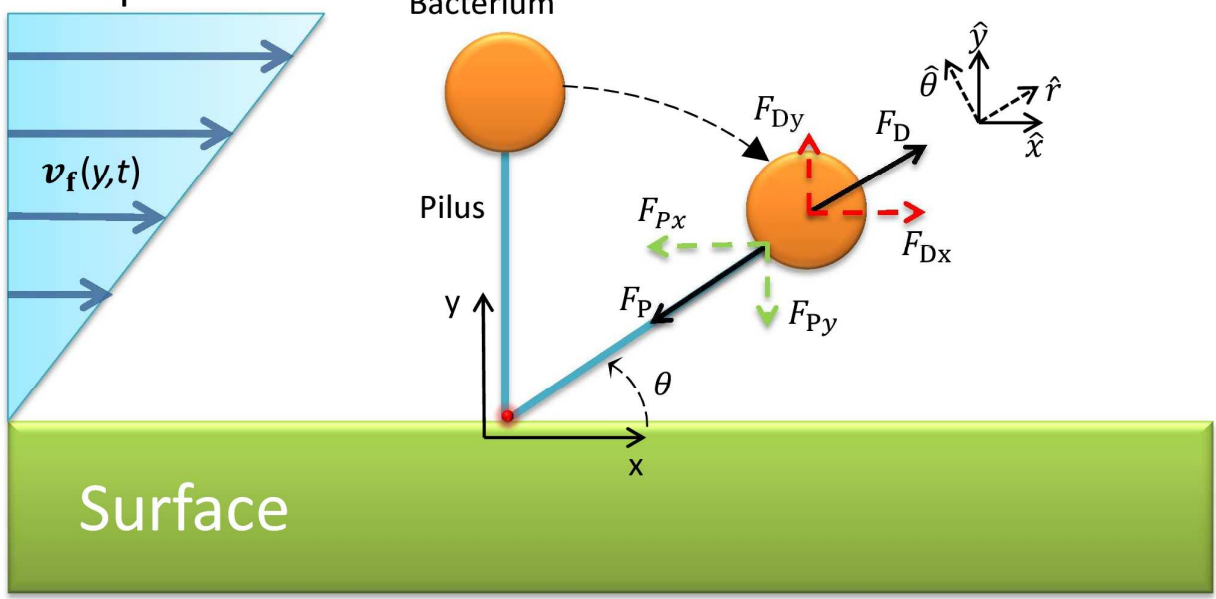

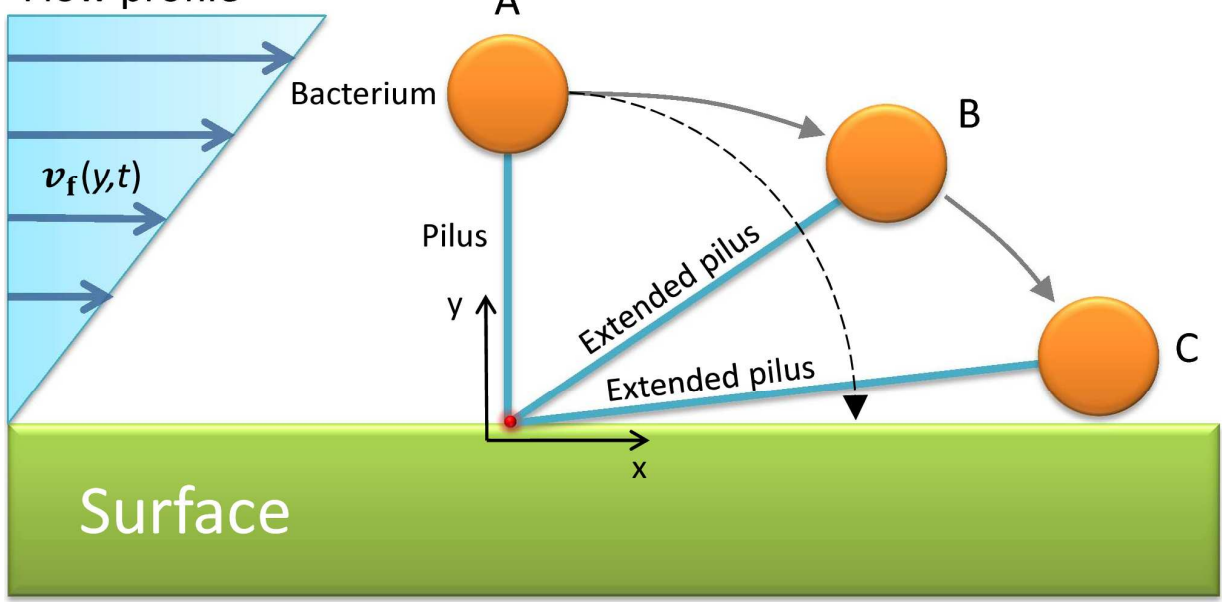

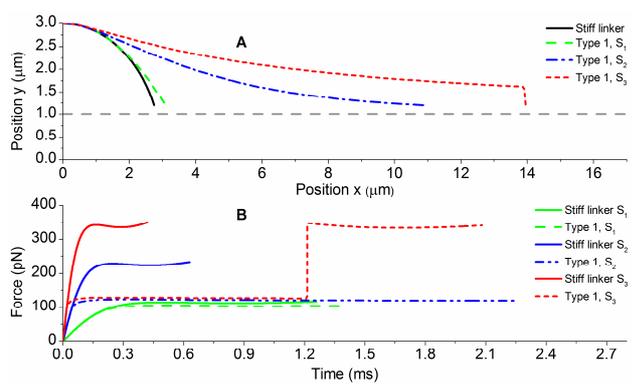

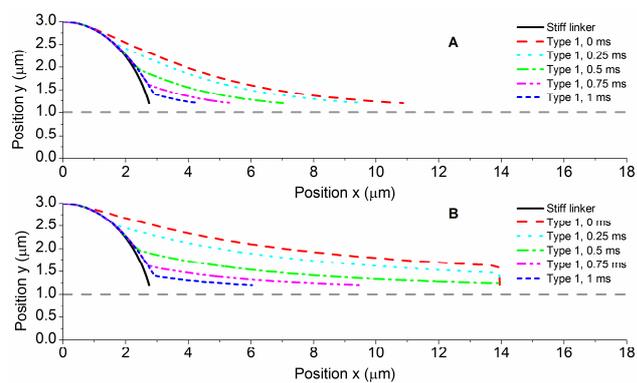

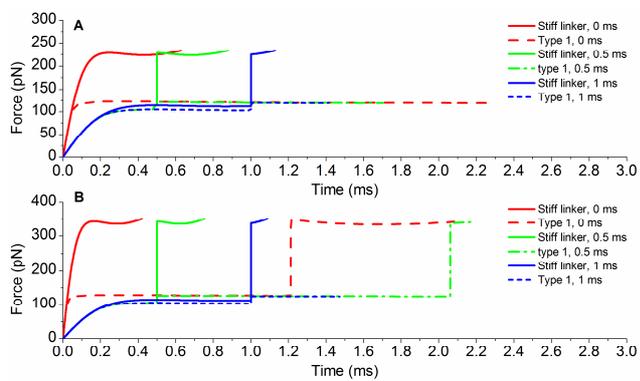

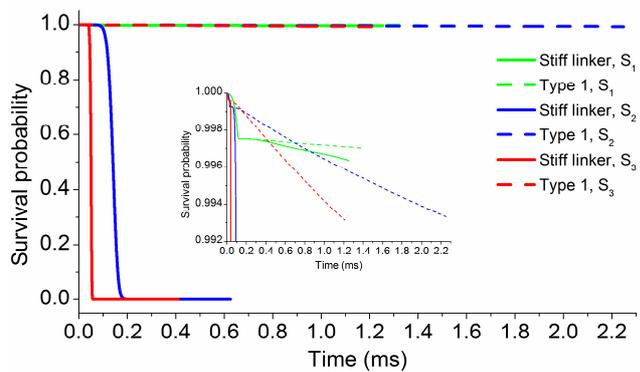